\documentclass[aps,prd,reprint,floatfix,nofootinbib,superscriptaddress,longbibliography]{revtex4-2}
\usepackage{amsmath,amssymb,bm,graphicx,booktabs}
\usepackage[colorlinks=true,linkcolor=blue,citecolor=blue,urlcolor=blue]{hyperref}
\newcommand{\DNSError}{5.41\times10^{-3}}
\newcommand{\DNSRadio}{3.38\times10^{-3}}
\newcommand{\DNSGW}{4.22\times10^{-3}}
\newcommand{\BHError}{1.02\times10^{-3}}
\newcommand{\BHRadio}{4.57\times10^{-4}}
\newcommand{\BHGW}{9.15\times10^{-4}}
\newcommand{\SkyRatioMin}{0.9767}
\newcommand{\SkyRatioMedian}{1.0000}
\newcommand{\SkyRatioMax}{1.0313}

\newcommand{\SkyChirpMax}{1.0002}

\newcommand{\eps}{\epsilon}
\newcommand{\Mc}{\mathcal{M}}
\newcommand{\dd}{\mathrm{d}}

\newcommand{\GW}{\mathrm{GW}}
\begin{document}
\title{The inclination cross-check: binary-pulsar timing and gravitational-wave polarimetry}
\author{Bhooshan Gadre}
\email{bhooshan.gadre@iucaa.in}
\affiliation{Inter-University Centre for Astronomy and Astrophysics (IUCAA), Post Bag 4, Ganeshkhind, Pune 411007, India}
\date{October 7, 2026}
\begin{abstract}
Radio timing and gravitational-wave (GW) polarimetry provide independent measurements of a binary's inclination. We study their consistency through $\eps=s_{\rm Shapiro}-s_{\rm GW}$, where $s_{\rm Shapiro}$ is the Shapiro-delay shape and $s_{\rm GW}$ is inferred from the tensor polarization amplitudes. We combine a time-dependent, finite-arm detector response with radio timing averaged over each integration, retaining the amplitude--inclination covariance and the nonlinear orientation posteriors. For hypothetical 15-minute binaries at 5 kpc and $65^\circ$ inclination, four years of GW data and a 125-h radio campaign give $\sigma_\eps=5.4\times10^{-3}$ for a double neutron star and $1.0\times10^{-3}$ for a pulsar--black-hole binary when two static timing harmonics are fitted. An apsidally evolving radio model with $e=0.01$ and a free apsidal rate gives $4.5\times10^{-3}$ and $0.94\times10^{-3}$, respectively, with the circular GW covariance held fixed. Changes in the relative tensor amplitudes can be absorbed by the GW fit as an apparent inclination shift and revealed by the radio measurement. An inaccurate detector response can produce a similar discrepancy even when its Fisher errors are nearly unchanged. These forecasts use the $n=2$ GW harmonic and assume a fixed orbital plane, sustained radio visibility, and the specified timing precision; additional harmonics can provide further orientation information.
\end{abstract}
\maketitle
\raggedbottom

\section{Introduction}
Binary pulsars test gravity through the consistency of orbital dynamics and pulse propagation. In a favorable system, several post-Keplerian parameters can be measured independently and compared with the predictions of a theory for a single pair of component masses \cite{DamourTaylor1992,Stairs2003,FreireWex2024}. The Double Pulsar provides a particularly precise example \cite{Kramer2021}. Orbital decay in asymmetric binaries also constrains additional radiative channels, including the dipole radiation predicted by many scalar-tensor theories \cite{Freire2012,Zhao2022}. These tests motivate a complementary question: does the orbital orientation inferred from pulse propagation agree with that inferred from the gravitational radiation?

The Shapiro delay constrains inclination through the shape of the pulse-propagation delay in the companion's gravitational field. Gravitational waves (GWs) encode the same inclination in their polarization amplitudes. For a common orbital plane and adequate signal models, the two measurements should agree at the same reference epoch. We ask how precisely this equality could be tested with low-frequency GWs and radio timing, fitting the two inclinations independently.

An independent inclination measurement is useful because a narrowband tensor fit can absorb changes in relative polarization amplitudes into the inferred amplitude and orientation. The radio measurement can reveal this degeneracy. It is exact for the four-coefficient model at a fixed carrier frequency; broadband evolution, additional harmonics, precession, and other external information can break it. The comparison also tests the consistency of the radio and GW signal models and complements the broader fundamental-physics programme of LISA \cite{Barausse2020}.

Independent geometric tests of Shapiro delay have a precedent in PSR J0437$-$4715, whose three-dimensional orbit was measured through timing geometry \cite{vanStraten2001}. Electromagnetic observations can also constrain the geometry of GW sources: Lagos et al.~\cite{Lagos2024} use resolved radio-jet imaging to constrain the emitted polarization orientation and test propagation birefringence. Here we compare the Shapiro-delay shape of a bound pulsar binary with the inclination inferred from GW tensor ellipticity. A joint fit imposing one inclination would assume the equality that we wish to test.

Galactic double neutron stars in the LISA band, LISA-assisted radio searches, and the LISA--SKA science case provide the astrophysical motivation \cite{Lau2020,Kyutoku2019,Thrane2020,Miao2021}. Recent inference studies include eccentricity, periastron advance, and frequency evolution in Galactic neutron-star and black-hole binaries \cite{Middleton2026}. The predicted populations depend on binary evolution and Galactic history \cite{McMillan2026}. Our forecasts are conditional on discovering and timing a suitable binary; estimating the number of such systems requires a population and selection model.

We first consider circular binaries with fixed orbital planes. We retain the amplitude--inclination covariance, account for the radio observing schedule, and propagate both orientation posteriors to $\eps$. We then examine how detector-response errors and changes in polarization bias the comparison. The GW calculation uses a rigid, equal-arm constellation with effective stationary noise. Eccentricity, precession, radio visibility, and foreground anisotropy set the limits of this approximation and are discussed separately.

\section{The observable and its interpretation}\label{sec:observable}
\subsection{Independent measurements of inclination}
Let $z=\cos\iota$ and $s=\sin\iota=\sqrt{1-z^2}$. At leading quadrupole order, a circular binary has
\begin{align}
 h_+(t)&=A_0(1+z^2)\cos\Phi(t),\\
 h_\times(t)&=2A_0z\sin\Phi(t),\label{eq:wave}\end{align}
where
\begin{equation}
 A_0=\frac{2(G\Mc)^{5/3}(\pi f)^{2/3}}{c^4d},\qquad
 \Mc=\frac{(m_pm_c)^{3/5}}{(m_p+m_c)^{1/5}}.
\end{equation}
The dominant frequency is $f=2/P_b$ at Newtonian order. The signed polarization ratio is
\begin{equation}
 R=\frac{2z}{1+z^2},\qquad
 z=\frac{R}{1+\sqrt{1-R^2}}.\label{eq:ratio}
\end{equation}
This form of the inverse remains regular near $R=0$. The sign of the cross amplitude depends on the polarization and phase conventions. Changing $z$ to $-z$ reverses the handedness but leaves $s$ unchanged, so we retain both signs of $z$ when constructing the posterior for $s$.

For a circular orbit, the propagation and Roemer delays are
\begin{equation}
 \Delta_S=-2r\ln(1-s\sin\phi),\qquad
 \Delta_R=x\sin\phi,\label{eq:radio}
\end{equation}
with $r=T_\odot m_c$ and $x=a_p\sin\iota/c$. Here $a_p$ is the pulsar's barycentric semimajor axis, $T_\odot=GM_\odot/c^3$, and masses multiplying $T_\odot$ are in solar units. We use these GR relations to set the injected parameters, then fit $r$, $x$, and $s$ independently. The Shapiro range is therefore not constrained by a companion mass inferred under GR.

We define
\begin{equation}
 \eps(t_0)=s_{\rm Shapiro}(t_0)-s_{\GW}(t_0).\label{eq:epsilon}
\end{equation}
We assume a constant source orientation. For an evolving orbital plane, both measurements must refer to the same retarded orbital epoch $t_0$, with propagation delays and orientation evolution included consistently. The GR relation concerns the underlying parameters; finite-data posterior means can differ, especially near a parameter boundary.

For locally Gaussian, independent estimates,
\begin{equation}
 \sigma_\eps^2=\sigma_{s,\rm R}^2+\sigma_{s,\rm G}^2.\label{eq:errors}
\end{equation}
A joint fit with uncertain shared parameters adds $-2\operatorname{Cov}(s_{\rm R},s_{\rm G})$. Our calculation conditions on the known sky position and the radio ephemeris, and checks the impact of freeing GW frequency and chirp. By contrast, if GR is imposed, the common-inclination precision satisfies
\begin{equation}
 (\sigma_\iota^{\rm joint})^{-2}
 =(\sigma_\iota^{\GW})^{-2}+z^2\sigma_{s,\rm R}^{-2}.
\end{equation}
This common-inclination uncertainty answers a different question from the error on the consistency statistic.

\subsection{What a nonzero residual would mean}
A nonzero $\eps$ could indicate an error in the timing or GW model, an instrumental or propagation effect, or a physical change in the relation between the two inferred inclinations. The statistic alone cannot distinguish these explanations, and some departures from GR leave it unchanged.

For example, a leading propagation modification of the form
\begin{equation}
 \Delta_S=-(1+\gamma_{\rm PPN})\frac{Gm_c}{c^3}\ln(1-s\sin\phi)
\end{equation}
can be absorbed into a freely fitted range without changing $s$ \cite{Will2014,DamourTaylor1992}. This example describes one possible modification of the delay, rather than a complete strong-field scalar-tensor timing model. Similarly, a common rescaling of the tensor amplitudes leaves Eq.~\eqref{eq:ratio} unchanged, and phase or dispersion changes need not alter the ratio if the signal remains coherently tracked. The test is sensitive to changes in the inferred geometry, rather than to every difference between the dynamics and radiation predicted by a theory.

\subsection{Electromagnetic anchor and a dynamical companion test}
Interpreting the radio shape as an independent inclination requires an adequate propagation model. A modification of the delay's functional form, the source structure, or the fitted orbit can bias $s$, even though a simple range rescaling does not. At leading order, the logarithmic Shapiro delay follows from propagation along the undeflected ray; deflection and retardation enter through additional terms.

For a sufficiently eccentric binary, a complementary, GR-dependent inclination can be obtained from the mass function. With $n_b=2\pi/P_b$, $T_M=GM/c^3$, and $T_c=Gm_c/c^3$,
\begin{equation}
 s_{\rm PK}=x n_b^{2/3}T_M^{2/3}/T_c,
\end{equation}
where $T_M$ and $T_c$ can be inferred from the GR periastron-advance and Einstein-delay relations \cite{DamourTaylor1992,FreireWex2024}. This supplies a complementary, theory-dependent check. Using it as a prior on $s_{\rm Shapiro}$ would change the null hypothesis, and a comparison of the two radio estimators must retain their shared timing covariance. The circular model considered here has neither a measurable Einstein delay nor a defined periapse angle, so we do not forecast this additional test.

\section{Gravitational-wave likelihood}\label{sec:gw}
\subsection{Response and normalization}
We use an equal-arm triangle of side $L=2.5\times10^9\,\mathrm{m}$ whose plane is inclined by $\pi/3$ to the ecliptic, following the mission baseline \cite{Colpi2024}. Its annual motion is specified by
\begin{equation}
 Q(t)=R_z(\alpha)R_y(\pi/3)R_z(-\alpha),\qquad
 \alpha=2\pi t/\mathrm{yr}.
\end{equation}
The triangle vertices in its reference plane are
\begin{equation}
 \bm v_j=\frac{L}{\sqrt3}
 [\cos(2\pi j/3),\sin(2\pi j/3),0],
\end{equation}
and $\bm r_j(t)=Q(t)\bm v_j$. We neglect arm flexing. The annual modulation of the response helps distinguish the tensor polarizations \cite{Cutler1998,CornishRubbo2003,Takahashi2002}.

In ecliptic Cartesian coordinates we use
\begin{align}
 \bm n&=(\cos\beta\cos\lambda,\cos\beta\sin\lambda,\sin\beta),\\
 \bm p&=(-\sin\lambda,\cos\lambda,0),\qquad \bm q=\bm n\times\bm p.
\end{align}
Thus $\bm p\times\bm q=\bm n$, $e^+=\bm p\bm p-\bm q\bm q$ and $e^\times=\bm p\bm q+\bm q\bm p$. Positive $\psi$ rotates the source basis to $\bm p_\psi=\bm p\cos\psi+\bm q\sin\psi$ and $\bm q_\psi=-\bm p\sin\psi+\bm q\cos\psi$. We take $X,Y,Z$ at vertices $j=0,1,2$, respectively, and order the outgoing arms as $\bm u_j=(\bm r_{j+1}-\bm r_j)/L$, $\bm v'_j=(\bm r_{j+2}-\bm r_j)/L$, with indices modulo three. The time origin is the observing-span midpoint and $\alpha(0)=0$.

For a source in direction $\bm n$, let $\bm u_j$ and $\bm v'_j$ be the two unit arms leaving vertex $j$. With $f_*=c/(2\pi L)$, the Michelson-normalized detector tensor is
\begin{align}
 D_j^{ab}&=\frac{e^{2\pi if\bm n\cdot\bm r_j/c}}{2}\bigl[u_j^au_j^b\mathcal T(\bm n\cdot\bm u_j,f)\nonumber\\
 &\hspace{17mm}-v_j^{\prime a}v_j^{\prime b}\mathcal T(\bm n\cdot\bm v'_j,f)\bigr],\label{eq:response}\\
 \mathcal T(\mu,f)&=\frac12\left[
 \operatorname{sinc}\!\left(\frac{f(1+\mu)}{2f_*}\right)
 e^{-if(3-\mu)/(2f_*)}\right.\nonumber\\
 &\hspace{14mm}\left.+\operatorname{sinc}\!\left(\frac{f(1-\mu)}{2f_*}\right)
 e^{-if(1-\mu)/(2f_*)}\right].\label{eq:transfer}
\end{align}
The source direction $\bm n$ is opposite to the propagation direction $\bm k$, so $\mu=\bm n\cdot\bm u=-\bm k\cdot\bm u$. We use $\operatorname{sinc}x=\sin x/x$. Equations~\eqref{eq:response}--\eqref{eq:transfer} follow the equal-arm Michelson response of Ref.~\cite{CornishRubbo2003}, including the spacecraft-position phase. In the long-wavelength limit, both this phase and $\mathcal T$ tend to unity. We remove the common guiding-center Doppler phase using the known sky position and divide out the common time-delay-interferometry transfer factor when defining the strain channels.

We form
\begin{equation}
 D_A=(2D_X-D_Y-D_Z)/3,\qquad D_E=(D_Z-D_Y)/\sqrt3.
\end{equation}
At low frequency these are independent channels of equal sensitivity \cite{Prince2002}. Contracting with the polarization tensors gives $F_I^a=D_I:e_a$. We use the tensor polarizations for the main forecast and introduce a massless breathing mode with $e_b=\bm p\bm p+\bm q\bm q$ with $\bm p,\bm q\perp\bm n$ \cite{TintoAlves2010}.

We adopt the four-year instrument-plus-confusion sensitivity $S_n=S_{\rm inst}+S_c$ of Ref.~\cite{Robson2019}. To combine it with the explicit response, we use an effective, diagonal channel PSD
\begin{equation}
 P_A=P_E=\mathcal R(f)S_n(f),\qquad
 \mathcal R(f)=\frac{3/10}{1+0.6(f/f_*)^2}.\label{eq:noise}
\end{equation}
This effective noise model restores the sky-averaged normalization without counting the response loss twice \cite{Babak2021}; it approximates the channel covariance of a full TDI calculation. At 2.222 mHz, the response averaged over our 256 sky and polarization configurations is $\langle\sum_I(|F_I^+|^2+|F_I^\times|^2)/2\rangle=0.298639$, compared with $\mathcal R=0.297579$. Their ratio, 1.00356, quantifies the accuracy of the sensitivity approximation. For the long-wavelength calculation we set $\mathcal R=3/10$ and $\mathcal T=1$. Both calculations use the same stationary, isotropic-equivalent foreground.

\subsection{Slowly evolving carrier and amplitude coordinates}
We take $f_b$ to be the radio mean-longitude frequency, giving a circular GW carrier $f=2f_b$. With apsidal evolution, the mean-longitude and anomalistic frequencies differ: $f_\lambda=f_r+\dot\omega/(2\pi)$, and the dominant co-rotating line lies at $2f_r+\dot\omega/\pi=2f_\lambda$. The finite-eccentricity radio calculation below uses $P_r=900$ s and compares its information with the circular GW covariance. A coherent eccentric analysis must instead use the appropriate mean-longitude carrier. The resulting apsidal displacement, 17--45 Fourier bins for our examples, is too large to treat as a small frequency offset.

The coherent carrier contains $f$, $\dot f$, and the known Doppler phase. Averaging its rapid oscillations gives the inner product of complex envelopes,
\begin{equation}
 (a|b)=\sum_{I=A,E}\int_{-T/2}^{T/2}
 \frac{\operatorname{Re}[a_I^*(t)b_I(t)]}{P_I}\,\dd t.
\end{equation}
The factor follows from the one-sided real-strain product $2\int h_1h_2/P_I\,\dd t$ and the carrier average $\langle\cos^2\Phi\rangle=1/2$. The approximation requires many carrier cycles and little change in the response and PSD across the narrow signal bandwidth. These conditions hold for the benchmarks.

For a fixed carrier the envelope is linear in four real coefficients:
\begin{equation}
 H_I=A_0(F_I^+c_++F_I^\times c_\times),
\end{equation}
where dimensionless coefficients are
\begin{align}
 c_+&=a[(1+z^2)\cos2\psi+2iz\sin2\psi]e^{i\varphi},\\
 c_\times&=a[(1+z^2)\sin2\psi-2iz\cos2\psi]e^{i\varphi}.\label{eq:coeff}
\end{align}
The scale $a$ is the fitted amplitude relative to the injected $A_0$ and is not fixed by the radio masses or distance. We define $\bm c=(\operatorname{Re}c_+,\operatorname{Im}c_+,\operatorname{Re}c_\times,\operatorname{Im}c_\times)$. The coefficient likelihood is exactly Gaussian within the stated response and noise model:
\begin{equation}
 -2\ln\mathcal L_{\GW}=(\bm c-\widehat{\bm c})^T
 G(\bm c-\widehat{\bm c})+\mathrm{const}.
\end{equation}
We evaluate the Gram matrix $G$ over four calendar years of continuous GW data and obtain the covariance by singular-value decomposition of the noise-weighted design matrix, retaining all singular values. For uniformly distributed gaps independent of the response, a retained fraction $D$ gives $G\rightarrow DG$ and $\sigma_{s,G}\rightarrow\sigma_{s,G}/\sqrt D$ at fixed PSD. With $D=0.75$ and unchanged radio errors, the static-model uncertainties become $5.93\times10^{-3}$ and $1.15\times10^{-3}$. This simple scaling assumes unchanged foreground noise and does not describe gaps concentrated at particular detector orientations. Numerical integration and convergence details are given in the accompanying reproducibility notes.

The Fisher parameters are $(\ln a,z,\psi,\varphi)$. To assess the effect of ephemeris uncertainty, we also allow frequency and chirp offsets $\delta fT$ and $\delta\dot fT^2$, whose envelope derivatives are $2\pi i(t/T)H$ and $\pi i(t/T)^2H$. The sky position remains fixed, so this calculation tests small ephemeris offsets rather than sky uncertainty or more general phase-model errors.

\subsection{An analytic polarimetric benchmark}\label{sec:ideal}
For comparison, consider an ideal instrument measuring both polarization amplitudes with equal independent variance $\sigma_A^2$. Writing $D=1+6z^2+z^4$ and $\rho^2=A_0^2D/\sigma_A^2$, the Fisher matrix for $(\ln A_0,z)$ is
\begin{equation}
 \Gamma=\frac{A_0^2}{\sigma_A^2}
 \begin{pmatrix}D&2z(z^2+3)\\2z(z^2+3)&4(1+z^2)\end{pmatrix}.
\end{equation}
Marginalizing the amplitude yields
\begin{equation}
 \sigma_z=\frac{D}{2\rho(1-z^2)},\qquad
 \sigma_\iota=\frac{D}{2\rho\sin^3\iota}.\label{eq:ideal}
\end{equation}
Although distance cancels from the noiseless polarization ratio, the fitted amplitude remains correlated with inclination. This degeneracy produces the face-on divergence in Eq.~\eqref{eq:ideal}. The expression provides a useful ideal-polarimeter benchmark for the full detector response. The effect of electromagnetic information on these correlations has been studied in Refs.~\cite{Shah2012,Shah2013}.

\section{Radio likelihood and observing assumptions}\label{sec:radio}
\subsection{Observing schedule and timing model}
We assume 250 radio observing sessions distributed over four years. Each session comprises forty consecutive 45-s integrations, giving $10^4$ pulse times of arrival (TOAs) and a total exposure of 125 h. The session midpoints are drawn uniformly from a visibility window covering 70\% of each year. This annual window represents a simple scheduling loss rather than the visibility of a particular telescope. We assume independent white timing errors of $1\,\mu\mathrm{s}$ per integration, a precision that must be achieved for the individual pulsar and observing system. The exact schedule and its construction are supplied with the reproducibility package.

For an integration centered on $t_k$ with width $\tau$, the model is the averaged timing residual,
\begin{equation}
 \overline\Delta_k=\frac1\tau
 \int_{t_k-\tau/2}^{t_k+\tau/2}\Delta(t)\,\dd t.\label{eq:boxcar}
\end{equation}
This approximation describes a small pulse-profile shift after coherent orbital demodulation, with the profile otherwise stationary. We average the delay over the changing orbital phase within each integration. The averaging is especially important at high inclination, where the Shapiro feature near superior conjunction is narrow.

The injected phase is
\begin{equation}
 \phi(t)=\phi_0+2\pi[f_b t+\tfrac12\dot f_b t^2],\qquad f_b=P_b^{-1},
\end{equation}
with the midpoint as $t=0$ and $\dot f_b=\dot f/2$. Both $f_b$ and $\dot f_b$ are subsequently fitted independently. The GR relation is used only for the injection.

Besides $s,r,x,\phi_0,f_b,\dot f_b$, we fit a constant, linear and quadratic spin-phase residual and two static second-harmonic coefficients. The latter span the first-order ELL1 eccentric Roemer terms \cite{Lange2001}. Allowing them to vary independently provides a conservative treatment of possible orbit-locked timing structure, including profile-dependent shifts. For a relativistically precessing eccentric orbit, an evolving eccentricity vector is more physically motivated; we examine this alternative below, following ELL1k \cite{Susobhanan2018}. At exact circularity the Einstein delay vanishes, and an additional first-harmonic term would be degenerate with the fitted orbital amplitudes.

Let $J$ denote the exposure-averaged design matrix for these eleven parameters. The timing Fisher matrix is $J^TC^{-1}J$, with white $C=\sigma_{\rm TOA}^2 I$ in the baseline. We include no external orbital priors. A separate experiment adds an epoch-correlated term,
\begin{equation}
 C=\sigma_{\rm TOA}^2 I+\sigma_J^2\,\operatorname{blockdiag}(\bm 1\bm 1^T),
\end{equation}
with $\sigma_J=1\,\mu\mathrm{s}$ and forty TOAs in each block. This increases $\sigma_s$ by less than $10^{-8}$ fractionally in both examples. Each session spans two orbits, so a common session offset has little overlap with the higher Shapiro harmonics that constrain the shape. Red noise, chromatic propagation and changes in the pulse profile require a more general treatment \cite{Cordes2010}.

\subsection{Static and rotating orbital harmonics}
Write $\kappa=e\cos\omega$, $\eta=e\sin\omega$, and $\lambda$ for mean longitude. To first order in eccentricity,
\begin{equation}
 \Delta_R=x\left[\sin\lambda+\frac{\kappa}{2}\sin2\lambda
 -\frac{\eta}{2}\cos2\lambda-\frac32\eta\right].
\end{equation}
For fixed $\omega$, the final term is absorbed by the fitted constant. It becomes time dependent when the eccentricity vector rotates and must then be retained. We linearize about $e=0$ in $(\kappa_0,\eta_0)$, including the first-order Shapiro terms, and use the GR apsidal rate for the injection. The rate is held fixed in this limit because it is unidentifiable at exact circularity. At finite eccentricity, we fit both the rate and the Einstein-delay amplitude.

For the finite-eccentricity comparison we use the leading DD delay structure \cite{DamourDeruelle1986},
\begin{align}
 q&=\sin\omega(\cos U-e)+\sqrt{1-e^2}\cos\omega\sin U,\\
 \Delta&=xq-2r\ln(1-e\cos U-sq)+\gamma_E\sin U.
\end{align}
Kepler's equation determines $U$, and $\omega=\omega_0+\dot\omega t$. The local fit has thirteen parameters: $s,r,x,M_0,f_r,\dot f_r,e,\omega_0,\dot\omega,\gamma_E$ and the three spin/offset coefficients. We retain the leading delays and secular rates rather than the full post-Newtonian DD model. Table~\ref{tab:ablation} compares the resulting uncertainties for the same observations.

\begin{table}[t]
\caption{Radio timing uncertainties for different nuisance models and integration widths, with identical TOA errors and total exposure. Entries are $10^3\sigma_s$. The rotating $e=0$ tangent conditions on the GR apsidal rate; finite-$e$ rows fit that rate independently.}\label{tab:ablation}
\begin{tabular}{lrr}\hline\hline
Timing model & DNS & PSR+BH\\\hline
No second harmonic, instantaneous & 1.495 & 0.2018\\
Static pair, instantaneous & 2.983 & 0.4025\\
Static pair, 45-s integrations & 3.384 & 0.4567\\
Rotating tangent, 45-s integrations & 1.621 & 0.2188\\
$e=0.01$, 13 parameters, 45 s & 1.614 & 0.2178\\
$e=0.1$, 13 parameters, 45 s & 1.191 & 0.1607\\\hline\hline
\end{tabular}
\end{table}

At $e=0.01$, fitting the apsidal rate gives uncertainties within 0.5\% of those obtained in the circular limit with the GR rate fixed. The gain over the static-harmonic model therefore persists when the rate is free. Increasing the eccentricity to 0.1 reduces the degeneracy between the shape and the nuisance parameters: the ratio of marginalized to shape-only error falls from about 7.2 to 5.3 in both systems, while the shape-only error changes by less than 1\%. The improvement comes mainly from this change in covariance.

Fitting the static harmonic pair approximately doubles the radio error. Finite integrations increase it by a further 13.4\%, compared with 8.4\% when the static pair is omitted. By comparison, removing the annual visibility gaps at fixed total exposure changes the static-model errors by only 0.091\% and 0.019\% for the alternative schedule considered here.

Allowing a secular change in $x$ increases the errors by less than $10^{-8}$ fractionally. Although the observing schedule is asymmetric, the $\dot x$ derivative is nearly orthogonal to the shape derivative after the other timing parameters have been projected out. Annual and semiannual timing terms, including linear drifts of the annual terms, have a similarly small effect for this schedule. The corresponding derivatives, correlations and numerical checks are given in the reproducibility notes. These additional terms describe local timing perturbations; modelling an evolving pulse profile or orbital plane requires a more complete treatment.

\subsection{Shapiro information and nuisance marginalization}
The instantaneous circular derivatives include
\begin{align}
 \partial_s\Delta&=\frac{2r\sin\phi}{1-s\sin\phi},&
 \partial_r\Delta&=-2\ln(1-s\sin\phi),\\
 \partial_x\Delta&=\sin\phi,&
 \partial_\phi\Delta&=x\cos\phi+\frac{2rs\cos\phi}{1-s\sin\phi}.
\end{align}
The phase, frequency and frequency-derivative terms follow by the chain rule, before averaging over each integration with Eq.~\eqref{eq:boxcar}. We obtain the error in $s$ from the full covariance matrix after marginalizing all nuisance coefficients.

This distinction is visible in the low-inclination expansion,
\begin{equation}
 \Delta_S=2rs\sin\phi+rs^2\sin^2\phi
 +\frac23rs^3\sin^3\phi+\cdots.
\end{equation}
The Roemer amplitude and phase absorb the first harmonic, while the fitted constant and second harmonic remove further low-order structure. The surviving constraint comes from the resolved Shapiro shape. This harmonic interpretation also motivates the orthometric parametrization used in pulsar timing \cite{FreireWex2010}.

For the radio posterior, we retain the exact logarithmic dependence on $s$ and analytically marginalize over the positive range $r$. We linearize the remaining nine timing parameters at the injection and project out their contributions. Thus the Shapiro likelihood is nonlinear in shape, while the nuisance timing model remains local. Comparison with a full timing package such as TEMPO2 or PINT \cite{Edwards2006,Luo2021} would be needed before applying this approximation to eccentric, strongly precessing or profile-variable systems.

\section{Posterior construction and validation}\label{sec:posterior}
The Gaussian coefficient likelihood can be transformed directly into a posterior for the physical parameters. It could also be used in a standard Bayesian sampler such as PyCBC Inference \cite{Biwer2019}.

Define the circular polarization coefficients
\begin{equation}
 c_R=c_++ic_\times,\qquad c_L=c_+-ic_\times.
\end{equation}
Equation~\eqref{eq:coeff} implies
\begin{align}
 c_R&=a(1+z)^2e^{i(\varphi+2\psi)},\\
 c_L&=a(1-z)^2e^{i(\varphi-2\psi)}.
\end{align}
With $u_R=\sqrt{|c_R|}$ and $u_L=\sqrt{|c_L|}$,
\begin{equation}
 z=\frac{u_R-u_L}{u_R+u_L},\qquad
 a=\frac{(u_R+u_L)^2}{4}.\label{eq:inverse}
\end{equation}
We draw samples from the Gaussian coefficient likelihood and transform them using Eq.~\eqref{eq:inverse}. The Jacobian is
\begin{equation}
 \left|\frac{\partial\bm c}{\partial(a,z,\psi,\varphi)}\right|
 =4a^3(1-z^2)^3.
\end{equation}
For a prior uniform in $z$, $\psi$, $\varphi$ and $\ln a$, the transformed samples therefore have weights proportional to $a^{-4}(1-z^2)^{-3}$. We use $0.1<a<10$; the posteriors lie well within these limits. On the radio side, the orientation prior is isotropic, or uniform in $|z|$, with a locally flat prior on the positive range $r$ and flat priors over the constrained linear nuisance parameters.

We first use noise-free injections to examine the likelihood geometry. The posterior for $\eps$ is the convolution of the independent radio and GW posteriors. Sample sizes, quadrature choices and convergence checks are documented in the reproducibility notes. Near face-on, the singular Jacobian requires particular care; the examples below do not establish reliable inference in that limit.

We then assess interval coverage with 1000 noise realizations for each source. We add Gaussian noise to the GW coefficients and draw the radio estimate from its local Gaussian sampling distribution. This tests the nonlinear GW transformation and the construction of the residual posterior, with the radio likelihood treated at Fisher order. Table~\ref{tab:coverage} gives the fraction of equal-tailed 68\% and 95\% intervals containing the injected value $\eps=0$. The corresponding binomial sampling uncertainties are about 0.015 and 0.007.

\begin{table}[t]
\caption{Reduced-likelihood residual-interval coverage in 1000 noise realizations per source. $p$ is the posterior mean of $\eps$ divided by its posterior standard deviation. The radio noise is Gaussian at the Fisher level in this experiment.}\label{tab:coverage}
\begin{tabular}{lrrrr}
\hline\hline
System & $C_{68}$ & $C_{95}$ & $\langle p\rangle$ & $\sigma_p$ \\
\hline
DNS & 0.71 & 0.95 & 0.08 & 0.97 \\
PSR+BH & 0.68 & 0.95 & -0.03 & 1.00 \\
\hline\hline
\end{tabular}

\end{table}

\begin{figure*}[!t]
\includegraphics[width=\textwidth]{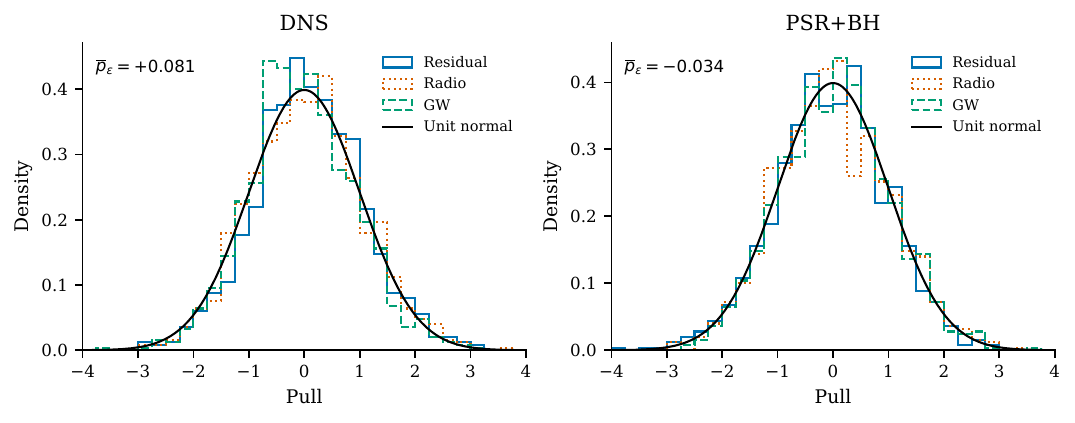}
\caption{Pull distributions from 1000 noise realizations for each source, using the static-harmonic radio model. The pull is the posterior mean minus the injected value, divided by the posterior standard deviation. Solid curves show the inclination residual; dotted and dashed curves show the separate radio and GW shapes. The black curve shows a unit normal distribution for comparison. Radio estimates are Gaussian at the local Fisher level.}\label{fig:pulls}
\end{figure*}

Figure~\ref{fig:pulls} shows the pull distributions. The mean radio and GW pulls are $+0.045$ and $-0.048$ for the DNS, and $+0.007$ and $+0.044$ for PSR+BH. Replacing the nonlinear GW posterior by its local Gaussian approximation changes the mean residual pull from 0.081 to 0.039 for DNS and from $-0.034$ to $-0.043$ for PSR+BH. The DNS offset therefore includes both finite-sample fluctuations and nonlinear GW inference. The isotropic radio prior and positive-range marginalization do not enter this experiment, since its radio estimator is Gaussian. The DNS 68\% coverage excess and mean pull lie about 2.2 and 2.6 sampling standard errors from their reference values. These tests use fixed injections; they are distinct from prior-predictive simulation-based calibration \cite{Talts2018}, which we have not performed near face-on.

We also check the analytic polarization covariance, the coefficient transformation, the response normalization, and the timing derivatives against the numerical implementation. Further checks cover exposure averaging, harmonic sums, noise whitening and residual projections. Their tolerances and convergence results are provided with the code. Such checks are necessary even at high SNR, where a narrow posterior can still be biased by an inadequate model \cite{Vallisneri2008}.

\section{Forecast results}\label{sec:results}
We consider two circular binaries, both with $P_b=900\,\mathrm{s}$, $d=5\,\mathrm{kpc}$, and $\iota=65^\circ$. The DNS masses are $(1.4,1.35)M_\odot$; the pulsar--black-hole masses are $(1.4,10)M_\odot$. The fiducial ecliptic longitude and latitude are $(60^\circ,30^\circ)$, the polarization angle is $0.4$ rad, and the GW initial phase is $0.3$ rad. The injected radio orbital phase is $0.2$ rad. A free GW phase allows these conventions to be related without imposing an unnecessary absolute phase constraint. The observing duration is four years.

Both binaries are hypothetical. Although DNS systems are observed, no assumption is made about the existence of a radio-visible 15-minute system at the chosen distance. The pulsar--black-hole example illustrates the sensitivity for a more massive companion. We assume fixed orbital planes; systems with appreciable spin misalignment require an evolving waveform and a separate assessment of radio visibility.

\begin{table}[t]
\caption{Forecast uncertainties for the fiducial binaries. S denotes the static-harmonic timing model and R the evolving $e=0.01$ radio model with a free apsidal rate. Both use the circular GW covariance; R is therefore a radio-model comparison rather than a joint eccentric forecast. BH denotes PSR+BH. Radio errors assume 125 h of timing with $1\,\mu\mathrm{s}$ TOAs, and GW errors use the finite-arm response. The two inclinations are fitted independently. Shape and residual errors are dimensionless.}\label{tab:benchmarks}
\begin{tabular}{lrrrr}
\hline\hline
System & $\rho$ & $\sigma_{s,\rm R}$ & $\sigma_{s,\rm G}$ & $\sigma_\epsilon$ \\
\hline
DNS (S) & 141.8 & $3.38\times10^{-3}$ & $4.22\times10^{-3}$ & $5.41\times10^{-3}$ \\
DNS (R) & 141.8 & $1.61\times10^{-3}$ & $4.22\times10^{-3}$ & $4.52\times10^{-3}$ \\
BH (S) & 653.7 & $4.57\times10^{-4}$ & $9.15\times10^{-4}$ & $1.02\times10^{-3}$ \\
BH (R) & 653.7 & $2.18\times10^{-4}$ & $9.15\times10^{-4}$ & $9.40\times10^{-4}$ \\
\hline\hline
\end{tabular}

\end{table}

For the DNS, the radio and GW errors $\sigma_{s,\rm R}=\DNSRadio$ and $\sigma_{s,\rm G}=\DNSGW$ give $\sigma_\eps=\DNSError$ (Table~\ref{tab:benchmarks}). The corresponding values for PSR+BH are $\BHRadio$, $\BHGW$ and $\BHError$. The GW inclination errors are $0.572^\circ$ and $0.124^\circ$. Radio timing contributes about 39\% and 20\% of the residual variance in the static-harmonic model. Substituting the $e=0.01$ radio errors reduces these fractions to 12.8\% and 5.4\%; at $e=0.1$ they are 7.4\% and 3.0\%. These comparisons hold the circular GW covariance fixed and show how the timing model affects the result.

The nonlinear residual standard deviations are $5.44\times10^{-3}$ and $1.023\times10^{-3}$, agreeing with the local Fisher values to within one percent. The DNS 95\% interval is approximately $[-1.04,1.10]\times10^{-2}$; the pulsar--black-hole interval is $[-1.98,2.03]\times10^{-3}$. The small median offsets arise from the nonlinear transformation and priors even when the injected parameters satisfy the null relation.

At fixed TOA midpoints and variance, averaging over the 45-s integrations raises the radio shape errors from $2.98\times10^{-3}$ to $3.38\times10^{-3}$ for DNS and from $4.02\times10^{-4}$ to $4.57\times10^{-4}$ for PSR+BH. This comparison measures the loss from smoothing the Shapiro feature. Shortening real integrations would also change their timing noise.

\begin{figure*}[!t]
\includegraphics[width=\textwidth]{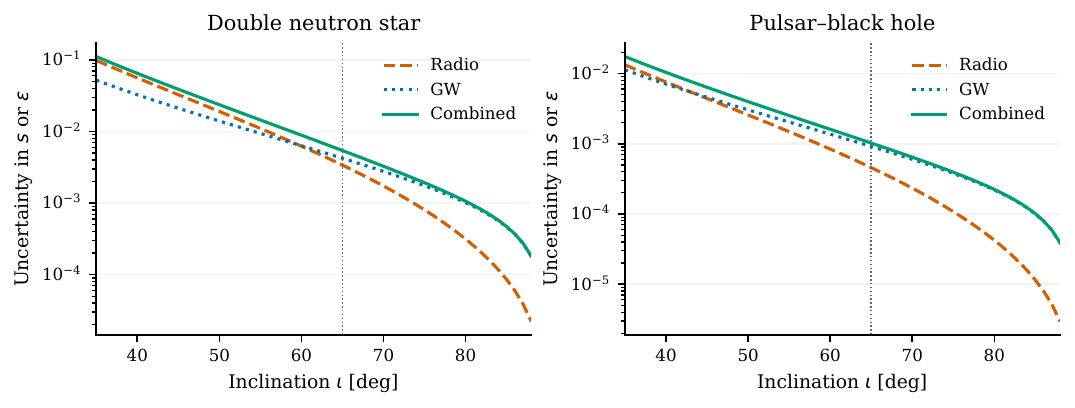}
\caption{Inclination dependence of the forecast errors for the two fiducial binaries, with static timing harmonics fitted. The radio (dashed), GW (dotted), and combined (solid) errors use the same source and observing assumptions. The calculation extends to $88^\circ$; the exactly edge-on posterior is discussed separately.}\label{fig:inclination}
\end{figure*}

Figure~\ref{fig:inclination} separates radio and GW information as inclination changes. At $88^\circ$, the DNS residual Fisher uncertainty is $1.81\times10^{-4}$ and the nonlinear posterior standard deviation is $1.83\times10^{-4}$. The posterior is visibly asymmetric even though the two width measures remain close. Exactly edge-on, first-order propagation gives zero GW uncertainty in $s$ and is invalid; Appendix~\ref{app:boundary} and Fig.~\ref{fig:diagnostics} give the finite posterior result.

\begin{figure}[t]
\includegraphics[width=\columnwidth]{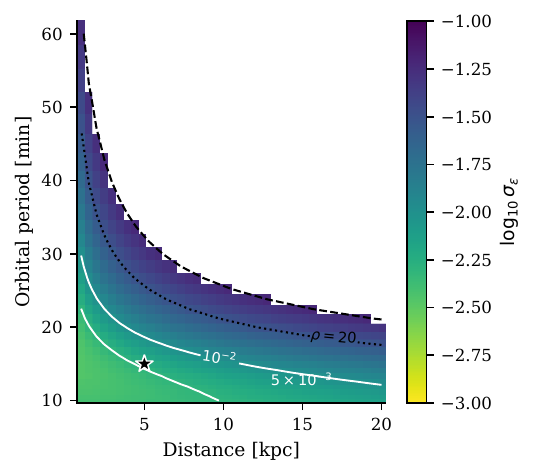}
\caption{DNS period--distance forecast with the static-harmonic radio model at $65^\circ$, with the radio schedule and per-TOA precision held fixed. Color gives $\log_{10}\sigma_\eps$. White regions have GW SNR below 10 and are excluded from the local forecast; the dashed curve marks this display cut and the dotted curve marks $\rho=20$. These SNR levels are used for display and do not define detection or parameter-estimation thresholds. The star is the fiducial binary. The radio precision is assumed to be maintained at every distance.}\label{fig:map}
\end{figure}

Figure~\ref{fig:map} extends the DNS calculation over orbital period and distance. Shorter periods increase the strain and move the signal towards the most sensitive part of the adopted noise curve. We show only points with $\rho\geq10$. The assumed radio precision is held fixed across the map; the observing time needed to achieve it is discussed in Sec.~\ref{sec:limits}.

To examine the variation with detector response, we sample 256 configurations uniformly in ecliptic longitude, sine latitude and $\psi\in[0,\pi)$, holding the masses, inclination and radio campaign fixed. The 5th--95th percentile range of $\sigma_\eps$ is $(5.31,5.77)\times10^{-3}$ for DNS and $(1.00,1.11)\times10^{-3}$ for PSR+BH. This ensemble describes variation over sky position and polarization rather than a Galactic source population.

\begin{figure*}[!t]
\includegraphics[width=\textwidth]{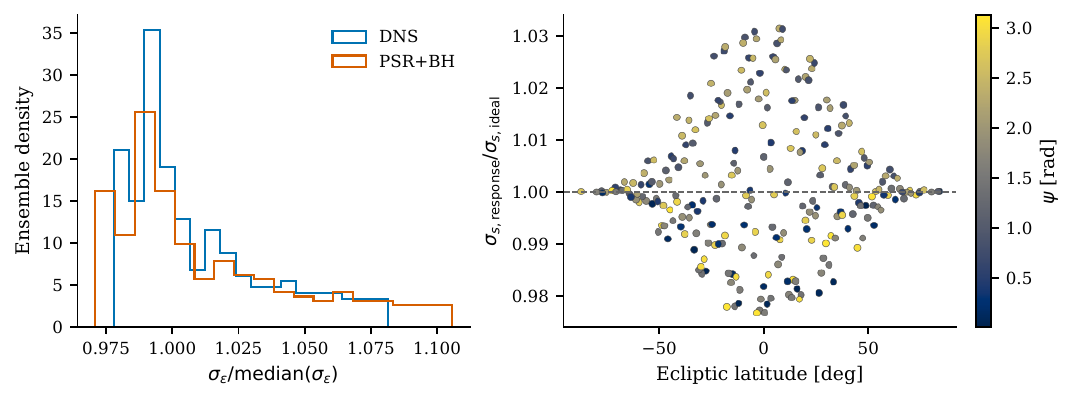}
\caption{Response variation in the 256-point isotropic sky/polarization ensemble. Left: the residual uncertainty divided by its ensemble median. Right: the GW shape error divided by the equal-polarization-noise estimate of Eq.~\eqref{eq:ideal}, using each source's calculated SNR. Point color indicates polarization angle; no Galactic-latitude foreground correction is applied.}\label{fig:sky}
\end{figure*}

Across this ensemble, the ratio of the GW shape error to the ideal-polarimeter estimate in Eq.~\eqref{eq:ideal} ranges from $\SkyRatioMin$ to $\SkyRatioMax$, with median $\SkyRatioMedian$ (Fig.~\ref{fig:sky}). Freeing frequency and chirp increases the GW shape error by at most a factor $\SkyChirpMax$. These small changes apply to the four-year, known-sky observation with stationary weighting considered here. Shorter observations, gaps or additional source parameters can introduce stronger correlations.

\begin{figure*}[!t]
\includegraphics[width=\textwidth]{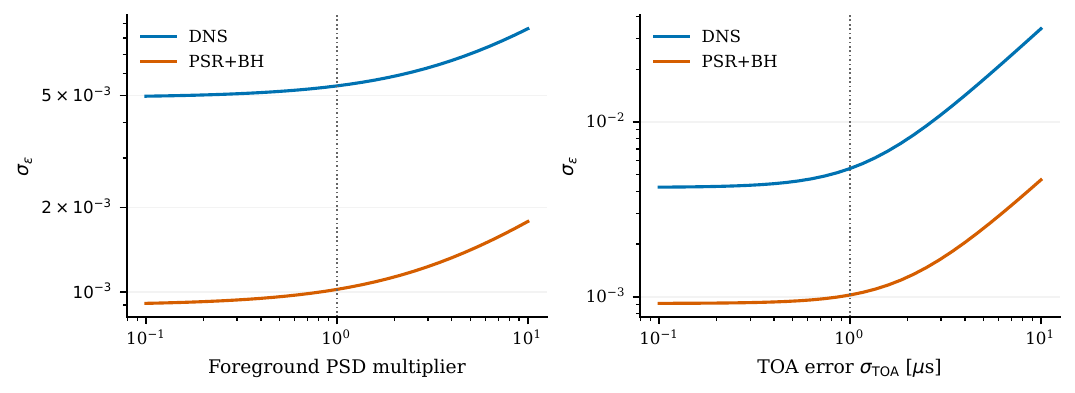}
\caption{Dependence on foreground level and timing precision, using the static-harmonic radio model. Left: a multiplier of the stationary confusion PSD at fixed instrument noise. Right: the per-TOA uncertainty at fixed cadence and exposure count. }\label{fig:noise}
\end{figure*}

Figure~\ref{fig:noise} shows how the result changes with foreground level and timing precision. Better timing improves the comparison until the radio error is small relative to the GW error. Varying the foreground amplitude estimates sensitivity to its overall level; an anisotropic treatment must also include the annual modulation of the channel covariance produced by unresolved Galactic sources \cite{Pozzoli2024}.

\section{Bias from response and polarization changes}\label{sec:deform}
\subsection{Parameter bias and residual power}
Let an injected signal be $h(\theta_0)+\delta h$, and fit it with $h(\theta)$. To first order,
\begin{equation}
 \delta\theta^a=(\Gamma^{-1})^{ab}(\partial_bh|\delta h),\label{eq:bias}
\end{equation}
while the component orthogonal to the model tangent remains in the residual \cite{CutlerVallisneri2007}. In the linear coefficient likelihood these projections are exact: if $B$ is the whitened design,
\begin{equation}
 \delta\bm c=(B^TB)^{-1}B^T\delta\bm d,\qquad
 \delta\bm d_\perp=\delta\bm d-B\delta\bm c.
\end{equation}
Transforming the shifted coefficients with Eq.~\eqref{eq:inverse} gives the nonlinear inferred orientation. The norm $\|\delta\bm d_\perp\|$ gives the residual SNR. Assigning a detection significance would require a separate model-selection calculation.

\subsection{Finite-arm signals fitted in the long-wavelength limit}
At the fiducial sky position, the finite-arm and long-wavelength calculations give Fisher errors differing by only 0.18\%. Their inferred inclinations can nevertheless differ substantially. Fitting a signal generated with Eq.~\eqref{eq:response} using the four-amplitude long-wavelength model gives
\begin{equation}
 \eps_{\rm LWL\ fit}\simeq-4.59\times10^{-3}
\end{equation}
for both masses at the same frequency and geometry. The common bias is independent of signal amplitude and corresponds locally to a polarization-ratio error of about $-1.6\%$, of order $(f/f_*)^2$. It is 0.85 and 4.5 times the respective residual uncertainties. The unabsorbed waveform residual has SNR 2.86 for DNS and 13.19 for PSR+BH.

The bias changes with sky position and polarization. Across the 256-point ensemble, its median is $9.8\times10^{-5}$ and its 5th--95th percentile range is $[-4.31,4.98]\times10^{-3}$. Dividing each absolute bias by that configuration's static-model uncertainty gives medians of $0.45\sigma_\eps$ and $2.36\sigma_\eps$, and 95th percentiles of $0.94\sigma_\eps$ and $4.90\sigma_\eps$. For PSR+BH, the bias exceeds $1\sigma_\eps$ in 81\% of the ensemble. Finite-arm modelling is therefore important for much of the high-SNR sample even at $f/f_*\simeq0.116$. Agreement between Fisher errors alone would not reveal this problem.

\subsection{Helicity-dependent amplification}\label{sec:birefringence}
The parameterized post-Einsteinian (ppE) framework describes departures from GR waveforms \cite{YunesPretorius2009}. In the present fit, a common amplitude rescaling is absorbed by the free amplitude, whereas a relative change between polarizations can shift the inferred inclination. Parity-violating Chern--Simons propagation provides one example through helicity-dependent amplification \cite{Alexander2008}. This effect is not generic to parity-even scalar-tensor gravity, so we treat the polarization gains as phenomenological parameters.

Amplitude birefringence gives unequal gains to the two circular tensor polarizations \cite{Okounkova2022,Ng2023,Lagos2024}. Within the observed narrow band, let $g_R=1+\delta_\beta/2$ and $g_L=1-\delta_\beta/2$, with $|\delta_\beta|<2$. Applying these positive gains to $c_R$ and $c_L$ and using Eq.~\eqref{eq:inverse} gives
\begin{equation}
 z_{\rm obs}=\tanh\!\left[\operatorname{atanh}z+
 \frac14\ln\frac{g_R}{g_L}\right].\label{eq:helicitymap}
\end{equation}
For small $\delta_\beta$ and an unchanged radio delay,
\begin{equation}
 \delta z=\frac{s^2}{4}\delta_\beta,
 \qquad \eps=\frac{zs}{4}\delta_\beta+O(\delta_\beta^2).
 \label{eq:bireps}
\end{equation}
The sign is set by the definition $\eps=s_{\rm radio}-s_{\rm GW}$ and our specified helicity labels. The linear response vanishes face-on and edge-on and is largest in magnitude at $45^\circ$ and $135^\circ$. Exactly edge-on the leading effect is quadratic, so the linear conversion to a gain uncertainty cannot be used there.

At $65^\circ$, a gain parameter $\delta_\beta=0.04$ gives an exact noiseless residual $3.860\times10^{-3}$, compared with $3.830\times10^{-3}$ at first order. This departure lies entirely within the four-coefficient tensor signal family: in this narrowband model it changes apparent amplitude and inclination without leaving an orthogonal waveform residual. An independently inferred inclination is therefore essential to this example.

The local uncertainty $4\sigma_\eps/(|z|s)$ in the gain parameter is 0.0565 for DNS and 0.0107 for PSR+BH. To relate this to propagation, consider $g_{R,L}=\exp(\pm\kappa d)$, so that $\delta_\beta\simeq2\kappa d$. At 5 kpc, the corresponding uncertainties in $\kappa$ are $5.65\times10^3$ and $1.07\times10^3\,\mathrm{Gpc}^{-1}$. A comparison with cosmological measurements requires the same propagation law and frequency dependence \cite{Ng2023}.

For example, adopting the linear-frequency law $\kappa(f)=\kappa_{100}(f/100\,\mathrm{Hz})$ gives $\sigma_{\kappa_{100}}\simeq2.54\times10^8$ and $4.80\times10^7\,\mathrm{Gpc}^{-1}$. A value $\kappa_{100}=1\,\mathrm{Gpc}^{-1}$ would produce only $\eps\simeq2.13\times10^{-11}$ at 5 kpc and 2.222 mHz. The proposed comparison therefore gives no competitive propagation bound for this frequency dependence. Relating a measured gain to a particular theory, including possible effects at the source, requires consistent calculations of GW generation, propagation and radio timing.

\subsection{Tensor ellipticity and a breathing component}
To examine a change in tensor ellipticity, we rescale the cross amplitude by $1+\delta_\times$ while leaving the plus amplitude and carrier unchanged. At fixed true inclination,
\begin{equation}
 \eps\simeq\frac{z^2(1+z^2)}{s^3}\delta_\times.
\end{equation}
At $65^\circ$, the coefficient is 0.28277. For PSR+BH, injecting $\delta_\times=0.02$ gives a recovered median $\eps=5.85\times10^{-3}$ and a 95\% interval $[3.74,8.00]\times10^{-3}$. This example quantifies sensitivity to a change in tensor ellipticity without assigning that change to a specific theory.

For a distinct response pattern, we inject a breathing envelope
\begin{equation}
 \delta H_I=bA_0s^2 F_I^b e^{i(\varphi+\varphi_b)},\qquad b=0.03,
\end{equation}
at the tensor carrier frequency, with a fixed phase offset $\varphi_b$. As this offset varies, $\eps$ spans approximately $[-6.05,6.10]\times10^{-4}$, less than the residual uncertainty in either example. The unabsorbed breathing signal nevertheless has SNR 1.37 for DNS and 6.31 for PSR+BH. Radiation at a different harmonic, such as a dipole component, would require a separate calculation.

\begin{figure*}[!t]
\includegraphics[width=\textwidth]{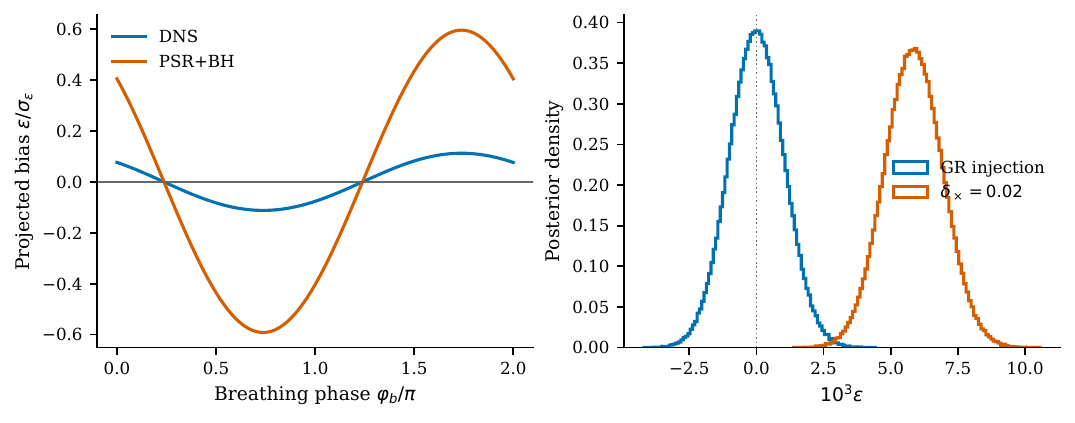}
\caption{Effects of changes in GW polarization. Left: inclination-residual bias from a phase-coherent breathing component of amplitude $b=0.03$, relative to the baseline uncertainty, as its phase is varied. The same physical shift has different statistical importance for the two masses. Right: pulsar--black-hole residual posteriors for the GR injection and a 2\% cross-amplitude deformation. The polarization changes are phenomenological.}\label{fig:deformations}
\end{figure*}

Fitting two independent breathing quadratures removes the noiseless inclination shift to numerical precision. Their marginalized errors are 0.02193 and 0.004757 for the two systems. The injected amplitudes have Gaussian Wald SNRs of 1.37 and 6.31, equal to the tensor-only residual SNRs as expected for a linear nested model. These are measures of signal strength, rather than Bayes factors or calibrated detection significances.

These examples show why both inclination and waveform residuals matter (Fig.~\ref{fig:deformations}). A departure absorbed by the tensor fit may appear as a disagreement in inclination, while a different polarization can leave appreciable residual power with little shift in $\eps$. Interpreting either observation within a gravity theory requires consistent timing and GW signals, including phase evolution and any dependence on stellar structure.

\section{Astrophysical reach and limits}\label{sec:limits}
\subsection{Source lifetime and sustained radio timing}
At the fiducial frequency, the circular chirps are $1.46\times10^{-16}$ and $6.74\times10^{-16}\,\mathrm{Hz\,s^{-1}}$. They move through 2.33 and 10.74 four-year Fourier bins. Both must be modeled coherently with a chirp even though their fractional frequency changes are small. Their circular coalescence times are $1.81\times10^5$ and $3.92\times10^4$ yr. A short remaining lifetime reduces the expected occupancy of a period interval, but does not by itself determine the number of radio-visible systems.

A population application would require, schematically,
\begin{align}
 N_{\rm useful}={}&\int \dd\vartheta\,\frac{\dd N}{\dd\vartheta}p_{\rm radio}(\vartheta)\nonumber\\
 &\times p_{\rm discovery}(\vartheta)p_{\rm timing}(\vartheta)\,\mathbf1_{\rm useful}(\vartheta).
\end{align}
where the factors account for beaming, radio activity, discovery, timing quality and GW selection. The LISA-assisted searches and population studies cited above address parts of this calculation; a sensitivity curve alone cannot determine the number of useful systems.

Achieving the assumed timing precision depends on the pulsar and telescope. In the radiometer-limited regime, $\mathrm{SNR}_{\rm pulse}\propto S_\nu\sqrt{\Delta\nu t_{\rm int}}$, and the TOA error scales roughly as $W_{\rm eff}/\mathrm{SNR}_{\rm pulse}$. For $W_{\rm eff}=1$ ms, a $1\,\mu$s error requires a folded-profile SNR of order $10^3$ in each 45-s integration, depending on the profile shape. Flux, width, scattering, jitter and calibration all affect whether this is attainable \cite{Cordes2010}. Maintaining the same precision at larger distances can demand substantially more telescope time.

For a fixed phase-sampling distribution and white noise, define $\mathcal I=N/\sigma_{\rm TOA}^2$. If a reference integration of duration $\tau_{\rm ref}$ achieves uncertainty $\sigma_{\rm ref}$, radiometer scaling gives
\begin{equation}
 T_{\rm exp}=\mathcal I\,\sigma_{\rm ref}^2\tau_{\rm ref}.
 \label{eq:timebudget}
\end{equation}
For $\mathcal I=10^{16}\,\mathrm{s}^{-2}$, a pulsar timed to $1$, $3$ or $10\,\mu$s in 60 s requires a total exposure of 167, 1500 or 16667 h, respectively. The 125-h campaign assumes $1\,\mu$s precision in 45 s. These estimates require white-noise averaging as $T_{\rm exp}^{-1/2}$ and refer to accumulated exposure rather than the elapsed observing span. Predicting $\sigma_{\rm ref}$ also requires the pulse profile, not just its flux and period.

\begin{figure*}[!t]
\includegraphics[width=\textwidth]{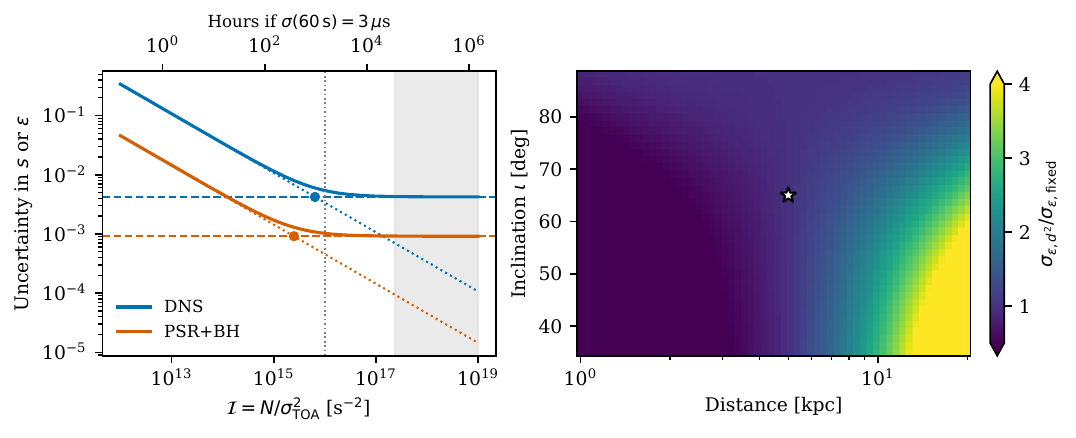}
\caption{Dependence on radio precision and observing time in the circular static-harmonic model. Left: combined residual error (solid), radio shape error (dotted), and GW floor (dashed); points mark equality of the two shape errors. The top axis converts timing precision to exposure using Eq.~\eqref{eq:timebudget}, with the phase sampling and integration smoothing held fixed. The shaded region would require more than four years of continuous observing at the reference precision, even before visibility losses and overheads. Right: DNS error under $\sigma_{\rm TOA}=1\,\mu\mathrm{s}(d/5\,\mathrm{kpc})^2$ divided by the fixed-$1\,\mu$s error, for $P_b=900$ s. The star marks the baseline. The curves use local Fisher scaling, whose accuracy has not been established in the large-error regime. }\label{fig:budget}
\end{figure*}

At fixed cadence, the radio and GW shape errors are equal when the per-TOA uncertainty reaches $1.25\,\mu$s for DNS and $2.00\,\mu$s for PSR+BH. At $30\,\mu$s, the residual errors rise to 0.102 and 0.0137. Figure~\ref{fig:budget} also considers $\sigma_{\rm TOA}\propto d^2$ for fixed pulsar luminosity, pulse shape and integration time. In this case the radio error grows faster with distance than the GW error, which scales as $d$, and eventually dominates.

The pulsar beam itself may evolve rapidly. The leading geodetic spin-precession rate is \cite{Barker1975,FreireWex2024}
\begin{equation}
 \Omega_p=T_\odot^{2/3}\left(\frac{2\pi}{P_b}\right)^{5/3}
 \frac{m_c(4m_p+3m_c)}{2(m_p+m_c)^{4/3}(1-e^2)}.
\end{equation}
For the circular examples, $\Omega_p=3.94$ and 16.16 rad yr$^{-1}$, corresponding to periods of 1.60 yr and 0.389 yr. This precession changes the spin direction rather than the orbital inclination. If the spin and orbit are misaligned, the pulse shape and beam visibility can evolve during the campaign. Nearly aligned systems are therefore the most favorable for sustained timing with a stationary profile, especially for a black-hole companion.

\subsection{Eccentricity and orbital-plane evolution}
Extending the circular forecast to eccentric binaries requires both messengers to be modelled consistently. The radio delay includes eccentric orbital motion and periastron advance; the GW signal includes multiple harmonics and precession sidebands. Relativistic periastron advance also separates the radial and azimuthal frequencies. Existing short-period timing and Galactic-binary inference studies provide starting points for such an analysis \cite{Miao2021,Middleton2026}.

The extra luminosity of an eccentric binary cannot be used as a direct multiplier of the circular strain SNR (Appendix~\ref{app:ecc}). Moreover, different harmonics carry different angular information, so even the total SNR is insufficient to determine the inclination covariance. The radio comparisons presented here retain the circular GW covariance; a joint eccentric forecast would have to calculate this harmonic information explicitly.

A J0737-like example with $e_0\simeq0.088$ at $P_{b0}\simeq2.45$ h gives $e\simeq e_0(P_b/P_{b0})^{19/18}\simeq0.008$ at 15 min under the small-$e$ Peters scaling. The $e=0.01$ diagnostic is therefore more representative of this formation path than $e=0.1$. At this eccentricity, the rotating $n=1,3$ strain-amplitude norms are approximately 0.00750 and 0.02250 relative to $n=2$, before detector weighting. Apsidal advance displaces the corresponding co-rotating lines by 17.47 and 45.08 Fourier bins relative to the anomalistic harmonics.

Misaligned companion spin can change the orbital orientation itself \cite{Wex1999,Kidder1995,Apostolatos1994}. In the orbit-averaged leading single-spin approximation, neglecting radiation reaction,
\begin{align}
 \dot{\bm L}&=K\bm S\times\bm L,&
 \dot{\bm S}&=-\dot{\bm L},\\
 K&=\frac{G(2+3m_p/2m_c)}{c^2a^3(1-e^2)^{3/2}}.
\end{align}
The total $\bm J=\bm L+\bm S$ is conserved. Both vectors precess around $\bm J$, with frequency $\Omega_J=KJ$, rather than around a fixed companion spin when $S\ll L$. If $\theta$ is the spin--orbit angle and $\lambda$ the angle between $\bm L$ and $\bm J$,
\begin{align}
 \sin\lambda&=\frac{S\sin\theta}{J},\\
 \cos\iota(t)&=\cos i_J\cos\lambda\nonumber\\
 &\quad+\sin i_J\sin\lambda\cos(\Omega_Jt+\alpha_0).
 \label{eq:precessioncone}
\end{align}
The finite-time change obeys
\begin{equation}
 |\Delta\widehat{\bm L}|=2\sin\lambda
 |\sin(\Omega_J\Delta t/2)|\leq2\sin\lambda.
\end{equation}
At short times, this reduces to $KS\sin\theta\,\Delta t$. The small precession cone and the angular frequency $KJ$ together give the torque rate $KS$; no further factor of $S/L$ is needed.

\begin{figure*}[!t]
\includegraphics[width=\textwidth]{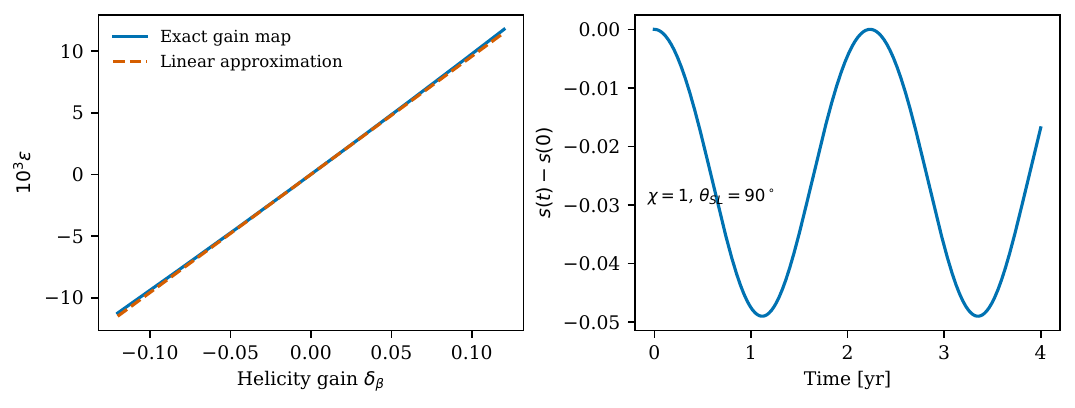}
\caption{Two distinct mechanisms affecting the comparison. Left: exact helicity-gain residual and its linear approximation at $65^\circ$. Right: a single-spin precession trajectory for the black-hole benchmark, with $\chi=1$, initial spin--orbit angle $90^\circ$, $\widehat{\bm L}(0)=(0,0,1)$, $\widehat{\bm S}(0)=(1,0,0)$, and line of sight $(\sin65^\circ,0,\cos65^\circ)$. The right panel shows the evolution of the common orbital plane; both messengers would follow it if modelled consistently.}\label{fig:interpretation}
\end{figure*}

For the circular black-hole example with $\chi=1$, $S/L=0.0523$ and $KS=0.1469\,\mathrm{rad\,yr^{-1}}$; the test-particle coefficient would give $0.1329\,\mathrm{rad\,yr^{-1}}$. At $\theta=90^\circ$, the cone angle is $\lambda=2.99^\circ$ and $\Omega_J=2.81\,\mathrm{rad\,yr^{-1}}$. Figure~\ref{fig:interpretation} shows the resulting inclination evolution for one viewing direction. Even this bounded motion can exceed the statistical error. A secular $\dot x$ alone cannot generally describe it over an appreciable fraction of the precession cycle.

For a small cone, the full-cycle excursion in $s$ is approximately $2|\cos i_J|\lambda$. Requiring it to remain below $10^{-3}$ at $i_J\simeq65^\circ$ gives $\lambda\lesssim1.18\times10^{-3}$ rad. On the nearly aligned prograde branch, $\lambda\simeq[S/(L+S)]\theta$; for $\chi=1$, this requires $\theta\lesssim1.4^\circ$. This sufficient condition depends on spin and viewing geometry. Slowly spinning or nearly aligned or antialigned systems can satisfy the fixed-plane approximation; more general systems require precessing inference.

\subsection{Population scale and omitted corrections}
A residence-time estimate gives an indication of how many very short-period binaries might be present. For a stationary flow of circular DNS binaries with the benchmark masses, the Galactic merger rate $37^{+24}_{-11}\,\mathrm{Myr}^{-1}$ \cite{Pol2020} and $t_c=0.181$ Myr give $N(P_b<900\,\mathrm{s})\simeq6.69$, with a rate-derived range of 4.70--11.0. The MMGPS-L update, $25^{+19}_{-9}\,\mathrm{Myr}^{-1}$ at 90\% credibility \cite{Colom2023}, gives $N\simeq4.52$ and a range of 2.89--7.95 under the same assumptions. These estimates precede all radio and GW selection effects. For NS--BH binaries, converting a volumetric merger rate into a useful Galactic population also requires a Galactic normalization and radio-active fraction. Small expected counts still allow a detection: a Poisson mean of 0.5, for example, gives a 39\% probability of at least one system.

We estimate the effect of two omitted signal terms by projecting them through the fitted model. For the moving-companion correction to the circular delay, let $A=1-s\sin\phi$ and $\beta_c=v_c/c$. Here $v_c$ is the magnitude of the companion's barycentric orbital velocity, so $\beta_c=(2\pi T_\odot M/P_b)^{1/3}m_p/M$ for masses in solar units. The orbital sign is carried by $\cos\phi$. Expanding the velocity-dependent propagation delay \cite{KopeikinSchafer1999,Kopeikin2002} gives
\begin{equation}
 \delta\Delta=2r\beta_c s\cos\phi
 \left[\ln A+\frac{s\sin\phi}{A}\right].\label{eq:moving}
\end{equation}
After exposure averaging and projection through the timing model, the shape biases are $3.25\times10^{-6}$ and $1.25\times10^{-6}$, with residual SNRs of 1.06 and 3.04. The bias in $\eps$ is small even though a timing residual remains. This calculation includes the first-order circular propagation correction. Emission aberration is a separate effect, with scale $(v_p/c)P_{\rm spin}/(2\pi)$, which can reach microseconds for $P_{\rm spin}=20$ ms. Its stationary first harmonic is absorbed by the orbital amplitude and phase, while evolution under geodetic precession requires a pulse-profile model.

The 1PN correction to the tensor amplitudes at the retained carrier \cite{Blanchet1996}, projected through the exact polarization inverse, gives $\eps=2.80\times10^{-7}$ and $1.96\times10^{-6}$, negligible compared with the benchmark errors. The derivation companion gives the separate plus/cross coefficients and their normalization. The leading odd harmonics at $f_b$ and $3f_b$ lie outside the retained carrier band and are orthogonal to it in the carrier-averaged approximation. Their relative amplitudes enter at order $v/c$ with a mass-asymmetry factor, so they can carry measurable SNR in the high-SNR PSR+BH case. Including them can add inclination information and break fixed-carrier degeneracies. A finite observing window or gaps can introduce leakage between these bands. In eccentric inference these frequencies overlap eccentric harmonics and must be modelled together.

\subsection{Modelling requirements}
An analysis of an observed system will require a fuller detector and timing model. In particular, the stationary foreground used here approximates a modulated covariance produced by unresolved Galactic sources \cite{Pozzoli2024}. A global foreground fit, flexing-arm TDI, clock and ephemeris uncertainties, and profile evolution may all affect the inferred polarization or timing shape. Eccentricity and precession must also be included when relevant to the source.

The effect of each correction is determined by its projection onto the fitted parameters. For example, arm-length variations of order one percent need not produce an equally large polarization-ratio error because a common response change can be absorbed by the amplitude. Quantifying the remaining bias requires injections with a flexing-arm TDI response. More generally, Eq.~\eqref{eq:bias} or a suitable injection study should establish $|\delta\eps|\ll\sigma_\eps$ for the source under consideration. As the finite-arm example shows, agreement in SNR or Fisher precision is not sufficient.

\section{Conclusions}
Radio timing and low-frequency GW polarimetry offer two measurements of the same orbital inclination. Their comparison provides a consistency test if the Shapiro range and shape are fitted independently and the GW inclination is inferred without a radio inclination prior. We have calculated its precision using a time-dependent finite-arm response, finite radio integrations and explicit timing nuisance parameters.

For the hypothetical 15-minute binaries considered here, four years of GW data and a 125-h radio campaign give $\sigma_\eps=5.4\times10^{-3}$ for DNS and $1.0\times10^{-3}$ for PSR+BH with the static timing harmonics fitted. The nonlinear posteriors agree closely with these local estimates. Using the evolving $e=0.01$ radio model with a free apsidal rate gives $4.5\times10^{-3}$ and $0.94\times10^{-3}$ at the same circular GW covariance. The two timing models answer different questions: the static pair allows unknown orbit-locked structure, while the rotating model describes an eccentric orbit more physically. Neither calculation includes the full eccentric GW signal, and both retain only the $n=2$ carrier. The stronger PSR+BH constraint also assumes little orbital-plane evolution, favoring small spin or near alignment, and remains conditional on discovering a suitable radio-visible system.

The comparison is particularly useful when a change in relative polarization amplitudes is absorbed into the apparent GW inclination. The helicity-gain example produces such a shift without an orthogonal tensor residual. Other effects leave residual power as well: the long-wavelength recovery of PSR+BH gives a false inclination discrepancy accompanied by residual SNR 13.19. Thus $\eps$ and the waveform residuals provide complementary information. Applying the test will require both radio and GW modelling errors to lie below the statistical precision, but it can then probe a consistency relation that is lost when the two inclinations are constrained to agree.

\section*{Data and code availability}
The code, simulated data, plotting scripts and manuscript sources accompany this paper. The package includes a derivation of the main results, the exact observing schedule, numerical settings, validation checks and instructions for reproducing the tables and figures. It uses no private observational data. Public archival deposition is pending.

\appendix
\section{Noise and numerical conventions}\label{app:noise}
The adopted four-year sensitivity is
\begin{align}
 S_{\rm inst}(f)&=\frac{10}{3L^2}\left[P_{\rm OMS}+\frac{4P_{\rm acc}}{(2\pi f)^4}\right]
 [1+0.6(f/f_*)^2],\\
 P_{\rm OMS}&=(1.5\times10^{-11}\,\mathrm m)^2
 [1+(2\,\mathrm{mHz}/f)^4]\,\mathrm{Hz}^{-1},\\
 P_{\rm acc}&=(3\times10^{-15}\,\mathrm{m\,s^{-2}})^2
 [1+(0.4\,\mathrm{mHz}/f)^2]\nonumber\\
 &\quad\times[1+(f/8\,\mathrm{mHz})^4]\,\mathrm{Hz}^{-1},\\
 S_c(f)&=9\times10^{-45} f^{-7/3}
 e^{-f^{0.138}-221f\sin(521f)}\nonumber\\
 &\quad\times[1+\tanh(1680(0.00113-f))]\,\mathrm{Hz}^{-1}.
\end{align}
The factor 4 in the acceleration term approximates $2[1+\cos^2(f/f_*)]$ at low frequency. At the fiducial frequency the exact factor is 0.99325 times this approximation, which we use throughout. Numerical frequencies in the last line are in hertz, as in the empirical fit \cite{Robson2019}. At 2.222 mHz, the instrument and confusion PSDs are $7.293\times10^{-40}$ and $2.878\times10^{-40}\,\mathrm{Hz}^{-1}$. The corresponding sky-averaged circular SNRs at $65^\circ$ are 139.5 and 643.4. The source-specific values in Table~\ref{tab:benchmarks} differ because the response is retained explicitly.

At low frequency, $\langle\sum_I |F_I^+|^2\rangle=\langle\sum_I |F_I^\times|^2\rangle=3/10$. Consequently
\begin{equation}
 \langle\rho^2\rangle=\frac{TA_0^2}{S_n(f)}(1+6z^2+z^4).
\end{equation}
The amplitude $A_0$ requires no further sky-averaging factor. We use $T_\odot=4.9254909476412675\,\mu\mathrm{s}$, a year of 365.25 days, and $1\,\mathrm{pc}=3.085677581491367\times10^{16}\,\mathrm m$. The reproducibility package records the numerical precision and integration settings.

\section{Circular Shapiro identity and the local likelihood}\label{app:radio}
The unmarginalized circular derivative norm obeys
\begin{align}
 I_s&=4r^2\left\langle\frac{\sin^2\phi}{(1-s\sin\phi)^2}\right\rangle\\
 &=\frac{4r^2}{s^2}\left[1-\frac2{\sqrt{1-s^2}}+
 \frac1{(1-s^2)^{3/2}}\right].
\end{align}
This follows by expanding $s^2\sin^2\phi=[1-(1-s\sin\phi)]^2$ and using the circular averages of $(1-s\sin\phi)^{-1}$ and $(1-s\sin\phi)^{-2}$. Although the derivative norm tends to $2r^2$ as $s\to0$, marginalization still removes the first-harmonic contribution that is degenerate with the Roemer delay.

For any nuisance block $a,b$, the usable local information is
\begin{equation}
 \Gamma_{ss}^{\rm marg}=\Gamma_{ss}-\Gamma_{sa}
 (\Gamma_{ab})^{-1}\Gamma_{bs}.
\end{equation}
In the nonlinear radio calculation, let $Q$ be an orthonormal basis for the whitened nuisance design and let $P_\perp=I-QQ^T$. Define $g(s)$ as the exposure-averaged, whitened unit-range Shapiro vector and $y$ as the projected data. With $a_s=\|P_\perp g(s)\|^2$ and $b_s=y^TP_\perp g(s)$, integrating $r>0$ gives
\begin{equation}
 \mathcal L(s)\propto a_s^{-1/2}
 \exp\left(\frac{b_s^2}{2a_s}\right)
 \Phi_N\left(\frac{b_s}{\sqrt{a_s}}\right),
\end{equation}
up to an $s$-independent factor; $\Phi_N$ is the standard normal cumulative distribution. The determinant factor is retained. The nuisance tangent is fixed at the injection, which is why this likelihood is described as local even though its Shapiro dependence is nonlinear.

\begin{figure*}[!t]
\includegraphics[width=\textwidth]{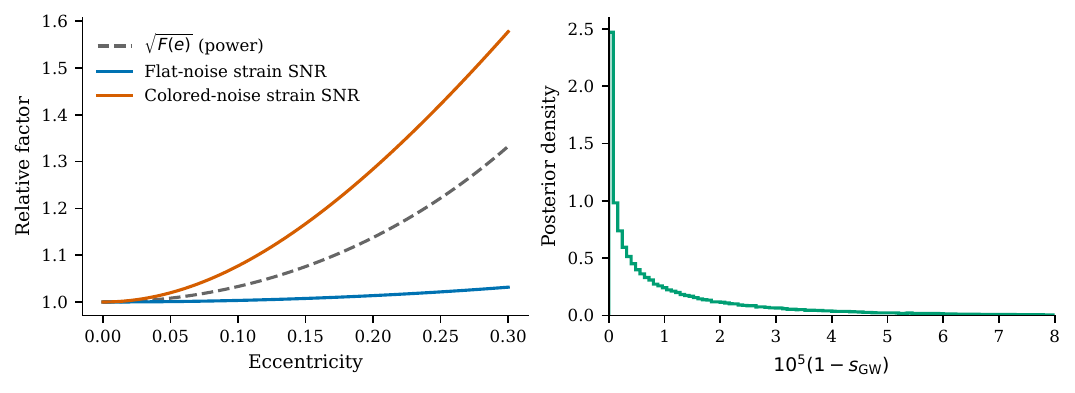}
\caption{Two analytic diagnostics. Left: eccentric luminosity and strain-SNR factors. The gray curve is the square root of the luminosity enhancement and is not a valid strain-SNR prescription. The blue and orange curves use the harmonic strain weights in flat and adopted colored noise, respectively, averaged over orientation. Right: the finite, asymmetric GW shape posterior for an exactly edge-on DNS. Linear propagation would incorrectly collapse this distribution to $s=1$.}\label{fig:diagnostics}
\end{figure*}

\section{The edge-on boundary}\label{app:boundary}
Near $z=0$,
\begin{equation}
 s=1-\frac{z^2}{2}+O(z^4).
\end{equation}
For a zero-centered Gaussian local distribution of $z$ with variance $\sigma_z^2$,
\begin{equation}
 E[s]\simeq1-\frac{\sigma_z^2}{2},\qquad
 \operatorname{Var}(s)\simeq\frac{\sigma_z^4}{2}.
\end{equation}
The uncertainty in $s$ is therefore finite and second order in $\sigma_z$, even though linear propagation would give zero. For an exactly edge-on DNS, the GW posterior has $\sigma_s=1.86\times10^{-5}$ and an asymmetric 95\% interval of approximately $[0.999934,0.999999987]$. The radio calculation at exact edge-on requires a separate treatment of companion geometry, visibility and the point-source propagation approximation \cite{RafikovLai2006}.

\section{Eccentric luminosity and strain signal-to-noise ratio}\label{app:ecc}
For a Newtonian eccentric orbit at fixed semimajor axis, the Peters--Mathews harmonic powers satisfy \cite{PetersMathews1963,Peters1964}
\begin{equation}
 \sum_{n\geq1}g(n,e)=F(e)
 =\frac{1+73e^2/24+37e^4/96}{(1-e^2)^{7/2}}.
\end{equation}
The orbit-averaged luminosity weights strain power by $f_n^2$, where $f_n=n/P_b$. The orientation-averaged squared strain SNR therefore satisfies
\begin{equation}
 \frac{\rho^2(e)}{\rho^2(0)}
 =\sum_{n\geq1}\frac{4g(n,e)}{n^2}
 \frac{S_n(2/P_b)}{S_n(n/P_b)}.\label{eq:ecc}
\end{equation}
This relation assumes narrow, slowly evolving harmonics and averages over sky position and orientation. It gives an SNR diagnostic rather than the inclination covariance of a particular source. In flat noise the harmonic sum is
\begin{equation}
 H(e)=\sum_n\frac{4g(n,e)}{n^2}
 =\frac{4/\sqrt{1-e^2}-1}{3}.
\end{equation}
At $e=0.1$, $F=1.06735$ but $H=1.00672$; at $e=0.3$, $F=1.77624$ but $H=1.06438$. With the adopted colored noise and $P_b=900$ s, Eq.~\eqref{eq:ecc} instead gives squared-SNR ratios 1.160 and 2.490. The difference comes from the frequencies at which the harmonics sample the noise curve; their contribution to inclination precision also depends on their angular structure.

The rotating quadrupole lines lie at $nf_r\pm\dot\omega/\pi$, while the nonrotating lines have a different inclination dependence \cite{BarackCutler2004}. Their joint orientation information must be calculated from a consistent harmonic waveform. We evaluate the Bessel-function expression for $g(n,e)$ through the first 100 harmonics, which is converged over $0\leq e\leq0.3$, and check both the luminosity and flat-noise strain identities. These harmonic sums illustrate the distinction between radiated power and measured strain; they are not used to rescale the circular inclination forecast.

\section{Acknowledgement}
I thank Prof. Sanjit Mitra and Dr. Apratim Ganguly for helpful discussions. Generative artificial-intelligence tools were used during manuscript preparation, code development, and checks of the calculations.

\bibliographystyle{apsrev4-2}
\bibliography{references}
\end{document}